\documentclass[aps,prl,reprint,groupedaddress,superscriptaddress]{revtex4-1}
\usepackage{amsmath}
\usepackage{amssymb}
\usepackage{graphicx}
\usepackage{bm}
\usepackage{epstopdf}
\usepackage{color}
\usepackage{ulem}
\usepackage{endnotes}
\usepackage{eucal}
\usepackage[dvipsnames]{xcolor}
\usepackage{multirow}

\usepackage{float}
\usepackage{array}
\usepackage{booktabs}

\usepackage{graphicx}
\usepackage[caption=false]{subfig}
\usepackage{color}
\usepackage{amsfonts}
\usepackage{lipsum}
\usepackage[super]{nth}

\newcommand{\bv}[1]{\mathbf{#1}}
\newcommand{\ket}[1]{\left \vert #1 \right \rangle}

\newcommand{\ignore}[1]{}
\newcommand{\cnuc}{^{13}\text{C}}

\begin{document}

\title{Structural analysis of nuclear spin clusters \\ via two-dimensional nanoscale nuclear magnetic resonance spectroscopy}

\author{Zhiping Yang}
\affiliation{Hefei National Laboratory for Physical Sciences at the Microscale and Department of Modern Physics, University of Science and Technology of China, Hefei 230026, China}
\affiliation{CAS Key Laboratory of Microscale Magnetic Resonance, University of Science and Technology of China, Hefei 230026, China}

\author{Xi Kong}
\email{kongxi@nju.edu.cn}
\affiliation{The State Key Laboratory of Solid State Microstructures and Department of Physics, Nanjing University, Nanjing 210093, China}

\author{Zhijie Li}
\affiliation{Hefei National Laboratory for Physical Sciences at the Microscale and Department of Modern Physics, University of Science and Technology of China, Hefei 230026, China}
\affiliation{CAS Key Laboratory of Microscale Magnetic Resonance, University of Science and Technology of China, Hefei 230026, China}

\author{Kai Yang}
\affiliation{Hefei National Laboratory for Physical Sciences at the Microscale and Department of Modern Physics, University of Science and Technology of China, Hefei 230026, China}
\affiliation{CAS Key Laboratory of Microscale Magnetic Resonance, University of Science and Technology of China, Hefei 230026, China}

\author{Pei Yu}
\affiliation{Hefei National Laboratory for Physical Sciences at the Microscale and Department of Modern Physics, University of Science and Technology of China, Hefei 230026, China}
\affiliation{CAS Key Laboratory of Microscale Magnetic Resonance, University of Science and Technology of China, Hefei 230026, China}

\author{Pengfei Wang}
\affiliation{Hefei National Laboratory for Physical Sciences at the Microscale and Department of Modern Physics, University of Science and Technology of China, Hefei 230026, China}
\affiliation{CAS Key Laboratory of Microscale Magnetic Resonance, University of Science and Technology of China, Hefei 230026, China}
\affiliation{Synergetic Innovation Center of Quantum Information and Quantum Physics, University of Science and Technology of China, Hefei 230026, China}

\author{Ya Wang}
\affiliation{Hefei National Laboratory for Physical Sciences at the Microscale and Department of Modern Physics, University of Science and Technology of China, Hefei 230026, China}
\affiliation{CAS Key Laboratory of Microscale Magnetic Resonance, University of Science and Technology of China, Hefei 230026, China}
\affiliation{Synergetic Innovation Center of Quantum Information and Quantum Physics, University of Science and Technology of China, Hefei 230026, China}

\author{Xi Qin}
\affiliation{Hefei National Laboratory for Physical Sciences at the Microscale and Department of Modern Physics, University of Science and Technology of China, Hefei 230026, China}
\affiliation{CAS Key Laboratory of Microscale Magnetic Resonance, University of Science and Technology of China, Hefei 230026, China}
\affiliation{Synergetic Innovation Center of Quantum Information and Quantum Physics, University of Science and Technology of China, Hefei 230026, China}

\author{Xing Rong}
\affiliation{Hefei National Laboratory for Physical Sciences at the Microscale and Department of Modern Physics, University of Science and Technology of China, Hefei 230026, China}
\affiliation{CAS Key Laboratory of Microscale Magnetic Resonance, University of Science and Technology of China, Hefei 230026, China}
\affiliation{Synergetic Innovation Center of Quantum Information and Quantum Physics, University of Science and Technology of China, Hefei 230026, China}

\author{Chang-Kui Duan}
\affiliation{Hefei National Laboratory for Physical Sciences at the Microscale and Department of Modern Physics, University of Science and Technology of China, Hefei 230026, China}
\affiliation{CAS Key Laboratory of Microscale Magnetic Resonance, University of Science and Technology of China, Hefei 230026, China}
\affiliation{Synergetic Innovation Center of Quantum Information and Quantum Physics, University of Science and Technology of China, Hefei 230026, China}

\author{Fazhan Shi}
\email{fzshi@ustc.edu.cn}
\affiliation{Hefei National Laboratory for Physical Sciences at the Microscale and Department of Modern Physics, University of Science and Technology of China, Hefei 230026, China}
\affiliation{CAS Key Laboratory of Microscale Magnetic Resonance, University of Science and Technology of China, Hefei 230026, China}
\affiliation{Synergetic Innovation Center of Quantum Information and Quantum Physics, University of Science and Technology of China, Hefei 230026, China}

\author{Jiangfeng Du}
\email{djf@ustc.edu.cn}
\affiliation{Hefei National Laboratory for Physical Sciences at the Microscale and Department of Modern Physics, University of Science and Technology of China, Hefei 230026, China}
\affiliation{CAS Key Laboratory of Microscale Magnetic Resonance, University of Science and Technology of China, Hefei 230026, China}
\affiliation{Synergetic Innovation Center of Quantum Information and Quantum Physics, University of Science and Technology of China, Hefei 230026, China}

\begin{abstract}
Two-dimensional Nuclear Magnetic Resonance (NMR) is essential in molecular structure determination. The Nitrogen-Vacancy (NV) center in diamond has been proposed and developed as an outstanding quantum sensor to realize NMR in nanoscale. In this work, we develop a scheme for two-dimensional nanoscale NMR spectroscopy based on quantum controls on an NV center. We carry out a proof of principle experiment on a target of two coupled $\cnuc$ nuclear spins in diamond. A COSY-like sequences is used to acquire the data on time domain, which is then converted to frequency domain with the fast Fourier transform (FFT). With the two-dimensional NMR spectrum, the structure and location of the set of nuclear spin are resolved. This work marks a fundamental step towards resolving the structure of a single molecule.
\end{abstract}

\maketitle

Molecular structure analysis is a cornerstone of biology, chemistry and medicine. Among three vastly used techniques for structure analysis, X-ray \cite{algara-siller_square_2015}, electron microscopy \cite{adrian_cryo-electron_1984}, and nuclear magnetic resonance (NMR) \cite{aue_twodimensional_1976,oschkinat_three-dimensional_1988,wuthrich_way_2001}, NMR is the promising technique to reveal the structure information with nondestructive in vivo detection at room temperature. However, the conventional NMR relays on a large ensemble of molecules to obtain a sufficient good signal-to-noise ratio, which will average out some individual properties. NMR for single molecule structure analysis is of paramount importance, and a crucial step of this is single molecular two-dimensional NMR spectroscopy (2D NMR).

\begin{figure}
\centering
\includegraphics[width=1\columnwidth]{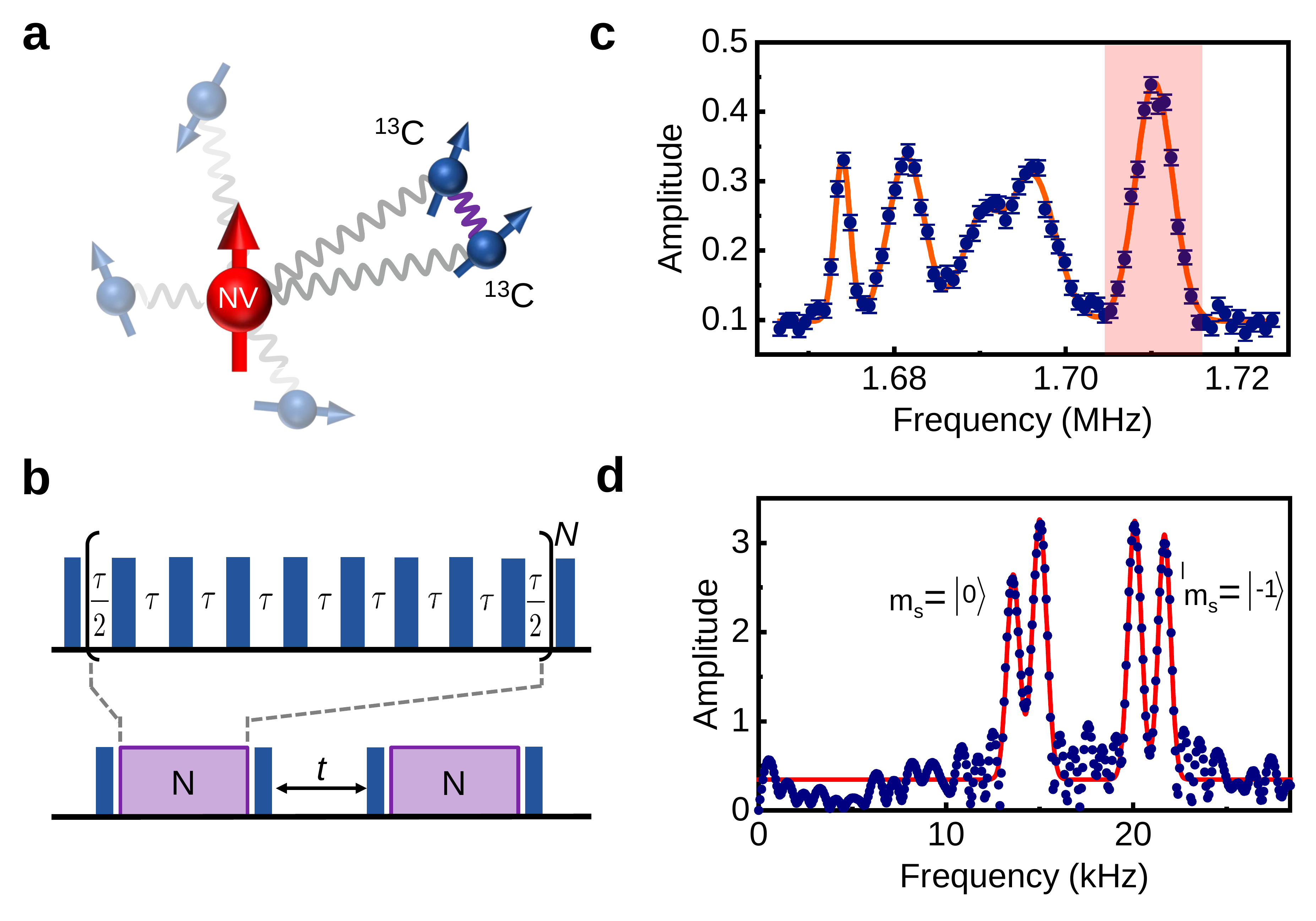}
\caption{Characteristics of the nuclear spin clusters with one-dimensional NMR.
(a) An [111]-orientated CVD-grown diamond is used in the experiment.  We study a NV-nuclear spins system while focus on two coupled  $^{13}$C nuclear spin cluster. (b) Pulse sequence. Upper, dynamical decoupling control sequence. Lower, correlation spectroscopy control sequence.   (c) Dynamical decoupling spectroscopy show the resonant frequency of the nuclear spins with $\omega_L-A_{\parallel}/2$. The fine spectrum is necessary to be investigated by high resolution correlation spectroscopy. We focus on the region labeled with red. (d) Fourier spectra of correlation spectroscopy on one of the $\cnuc$ spin cluster is observed with  180 mT magnetic field.
 }\label{fig:2d_theory}
\end{figure}

Thank to high sensitive atomical-scale NV centers \cite{balasubramanian_nanoscale_2008,maze_nanoscale_2008}, nanoscale magnetic resonance spectroscopy has been developed rapidly over the past years. One-dimensional nanoscale NMR \cite{staudacher_nuclear_2013,mamin_nanoscale_2013,lovchinsky_magnetic_2017}, single spin sensitivity NMR \cite{muller_nuclear_2014}, single molecule magnetic resonance \cite{shi_single-protein_2015,lovchinsky_nuclear_2016}, the detection of macroscopic-scale chemical shift and J-coupling \cite{aslam_nanoscale_2017,glenn_high-resolution_2018}, the 2D spectrum of NV-nuclear spin system \cite{boss_one-_2016,abobeih_atomic-scale_2019} and microscopic 2D NMR\cite{smits_two-dimensional_2019} have been realized with NV centers. While these works on nanoscale NMR make it possible to provide unprecedented insight into molecule structure, 2D nanoscale NMR \cite{ajoy_atomic-scale_2015,ma_proposal_2016} is critical for the unambiguous determination of molecule structure (Fig. \ref{fig:2d_theory}(a)).

In the present work, we propose a scheme for structure reconstruction of single molecule by 2D nanoscale NMR  based on the NV center in diamond. We carry out an experiment to characterize a pair of coupled $\cnuc$ nuclear spins in diamond and demonstrate experimentally 2D nanoscale NMR spectroscopy. The 2D information obtained offers high-confident analysis of three-dimensional molecular geometry structure. In the end, we evaluate the performance of our scheme on a typical single protein. 

We use a single NV center in a CVD-grown diamond with a natural abundance (1.1 \%) of $^{13}$C nuclear spins (Fig. \ref{fig:2d_theory}(a)). The NV electron spin is used as a quantum sensor to probe $^{13}$C nuclear spins cluster, with the pulse sequence plotted in Fig. \ref{fig:2d_theory}(b), through dynamical decoupling spectroscopy~\cite{du_preserving_2009,zhao_sensing_2012}. The NV electron spin is prepared in the superposition state $\ket{+}=(\ket{0}+\ket{-1})/\sqrt{2}$. During the controlled evolution, the coherence of NV electron spin decreases, leading to a dip on resonance by $\tau=1/2(\omega_L-A_{\parallel}/2)$, where $\omega_L$ is the Larmor frequency of nuclear spin and $A_{\parallel}$ is the parallel component of hyperfine interaction between NV electron spin and nuclear spin. The dips of the spectrum in Fig. \ref{fig:2d_theory}(c)  reveal nuclear spins with different hyperfine interactions. The high resolution correlation spectroscopy \cite{laraoui_high-resolution_2013,kong_towards_2015} is then applied to the NV and nuclear spins system to resolve the fine structure of spectrum by focusing on the red region in Fig. \ref{fig:2d_theory}(c) with center frequency $\omega_L=$1.71 MHz. Dynamical decoupling protocol with time interval resonant with $\omega_L$ is applied, and the correlation between the first and the second dynamical decoupling protocols is recorded. During the free evolution in between, the nuclear spins' evolution are dependent on the electron spin states $m_{\text s}=\ket{0},\ket{-1}$. Thus, the spectrum contains different peaks centered at $\omega_0=\omega_L$ and $\omega_{-1}=\omega_L-A_{\parallel}$. Actually, the spectrum  in Fig. \ref{fig:2d_theory}(d) shows two peaks for both $m_s=\ket{0}$ and $m_s=\ket{-1}$ subspaces.

However, such a spectrum in Fig. \ref{fig:2d_theory}(d) alone could allow different interpretations. It may correspond to a two coupling nuclear spin cluster or three isolated nuclear spins (see \cite{SM} section IIA). The one-dimensional NMR spectrum alone cannot discriminate these different interpretations. A two-dimensional correlation map is then necessary to provide such important information.

\begin{figure}
\centering
\includegraphics[width=1\columnwidth]{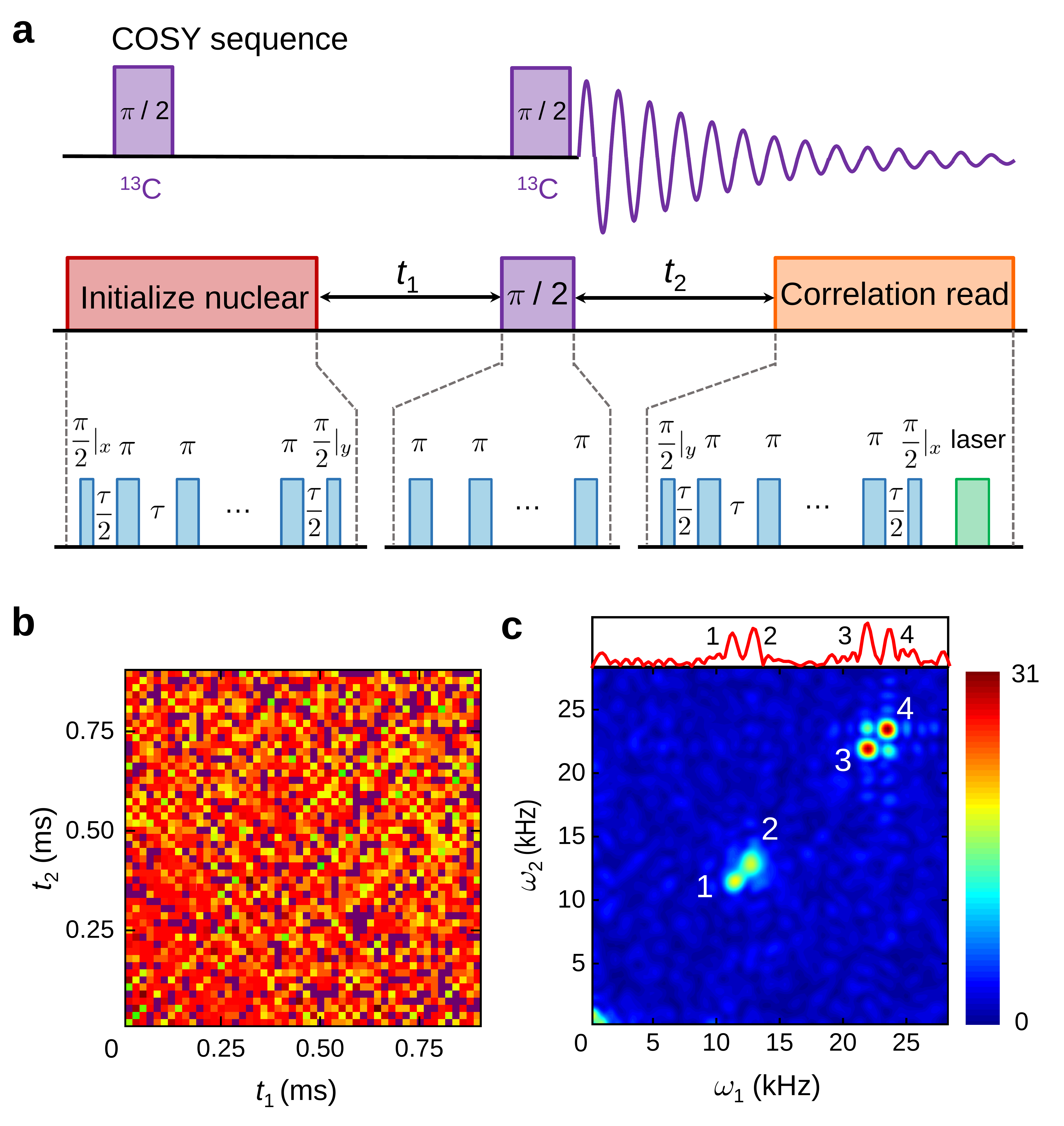}
\caption{Two-dimensional nuclear magnetic resonance and structure analysis on the coupled nuclear spin system.
(a) The experimental pulse protocol. The sequence is the experimental realization of the nanoscale homonuclear COSY two-dimensional nuclear magnetic resonance on NV center.  (b) We sweep $t_1$ and $t_2$ from 4 $\mu$s to 0.9 ms with 50 samplings. The time spectroscopy then converts to frequency spectroscopy in (c) by two-dimensional FFT transformation. (c) 2D NMR spectrum. There are cross peaks between the \nth{3} and \nth{4} peaks in one-dimensional spectrum upside, which indicate that they belong to a coupled spin system.
 }\label{fig:2d}
\end{figure}

In analog to the COSY spectroscopy in conventional NMR, we develop a two-dimensional protocol to investigate such correlation map of target nuclear spins (Fig. \ref{fig:2d}(a)).
The Hamiltonian for the system of an NV sensor and two nuclear spins is $ H= \omega_L (I_{1,z}+I_{2,z}) + \bv{S}\cdot(\bv{A}_{1}\cdot\bv{I}_1 + \bv{A}_{2}\cdot\bv{I}_2) + \bv{I}_1\cdot\bv{J}\cdot\bv{I}_2$, where $\bv{S}$ is the NV electron spin, $\bv{I}_i$ is the $i$th nuclear spin, $\bv{A}_i$ is the hyperfine coupling between NV and nuclear, $\bv{J}$ is dipolar coupling interaction between two nuclear spins. The NV sensor is driven with the dynamical decoupling sequence which is in resonant with the red region center frequency in Fig.~\ref{fig:2d_theory}(b). Then the target nuclear spin $\bv{I}_1$ is initialized to the transverse direction. During an interval of 0 to $t_1$, the two nuclear spins $\bv{I}_1$ and $\bv{I}_2$ interact and a phase $\phi_1=(\omega_L+A_{1,\parallel}S_z+J_{zz}\/m_{1}m_{2}) t_1$ is accumulated in the first nuclear spin, where $A_{1,\parallel}$ is the parallel component of the hyperfine interaction of the NV sensor with $\bv{I}_1$, and $J_{zz}$ is the $zz$ component of the coupling between nuclear spin $\bv{I}_1$ and $\bv{I}_2$. Then a series of $N$=40 dynamical decoupling $\pi$ pulse is applied, which corresponds to half $\pi$ pulse on the nuclear spin. A free evolution of duration $t_2$ comes afterwards, another phase $\phi_2=(\omega_L+A_{1,\parallel}S_z+J_{zz}\/m_{1}m_{2}) t_2$ accumulates. In the end, another dynamical decoupling sequence is applied to read out the transverse component of the nuclear spin. 

Sweeping the duration time $t_1$ and $t_2$ from 4 $\mu$s to 0.9 ms, the correlation map is realized, as shown in Fig. \ref{fig:2d}(b). The 2D FFT spectrum in Fig. \ref{fig:2d}(c) clearly shows the correlations of peaks of the one-dimensional spectrum, i.e. which peaks belong to the same spin system. In detail, the cross peaks between the \nth{3} and \nth{4} peaks indicate that they belong to a coupled spin system. The \nth{3} and \nth{4} peaks correspond to the condition $m_{\text s}=-1$, while the \nth{1} and \nth{2} peaks correspond to $m_{\text s}=0$.

\begin{figure}
\centering
\includegraphics[width=1\columnwidth]{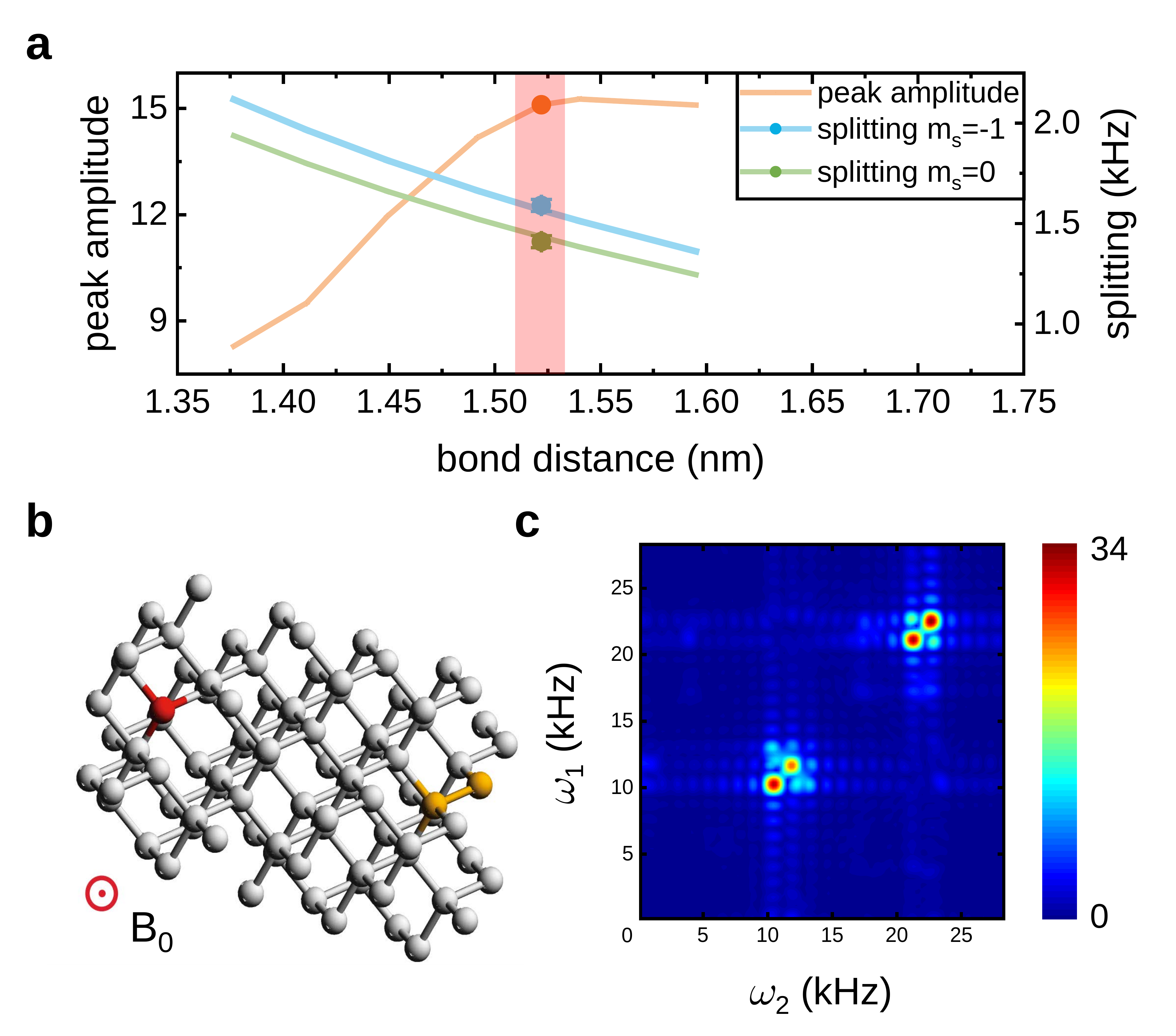}
\caption{Structure analysis on the coupled nuclear spin system.
(a) Estimation of the bond length of $^{13}$C. (b) The resolved position for the nuclear spin cluster. (c) The simulated spectrum according to the resolved nuclear spin cluster position.
}\label{fig:lattice}
\end{figure}

From the spectrum, we can obtain the bond length of off-axis $^{13}$C-$^{13}$C nuclear spins. The optimally estimated bond length is 1.52 $\pm$ 0.02 \AA ~(Fig. \ref{fig:lattice}(a)), (\cite{SM} section II D). We also resolve the position of the nuclear spin cluster in the diamond lattice as shown in Fig. \ref{fig:lattice}(b). The position is uniquely determined to be one of a set of symmetrically equivalent positions. Using these parameters, the simulated 2D spectrum is shown in Fig. \ref{fig:lattice}(b), which fits well with the experiment spectrum (minor difference in the \nth{1} and \nth{2} peaks due to field fluctuation).

The demonstrated two-dimensional NMR technique, with potential improvements in the sensor fabrication, may enable the capability to resolve the structure of a single molecule. We consider as an example the single avian pancreatic polypeptide molecular placed at a similar distance as the $\cnuc$ nuclear spin cluster to the NV center. Without loss of generality, two amino acid are labeled with $\cnuc$ and $^{15}\text{N}$ as shown in Fig. 4(a). To resolve the molecule structure, $\cnuc$ and $^{15}\text{N}$ are observed at the same time to get heteronuclear coupling. A non-periodical multi-nuclear species correlation sequence is introduced here (Fig. 4(b), \cite{SM} secrion III B). It consists of $\pi$ pulses inserted on zeros of $\cos(\omega_it)$ between two $\pi/2$ pulses to observe nuclear spins with all different frequency $\omega_i$. Unlike in an ensemble NMR experiment, the strong gradient field (about 0.1 mT/nm) from NV center~\cite{muller_nuclear_2014} splits the homonuclear spins spectrum of a single molecule. With an artificially fabricated antena~\cite{jakobi_measuring_2017}, a much larger gradient up to 10 mT/nm is possible. Utilized by multi-nuclear species correlation sequence, the 2D heteronuclear spin resonance sequence is shown in Fig. 4(d) in analog to the homonuclear one (Fig. 3(a)).  Then the simulated two-dimensional NMR spectrum is demonstrated in Fig. 4(d), the corresponding nuclear spins frequencies are labeled along the axis. The off-diagonal non-zero terms correspond to the coupling between different nuclear spins. By rotation the polar angle between the external magnetic field and the NV center, the dipolar interaction tensors are measured to characterize the distances between nuclear spins (\cite{SM} section III D). By solving the discretizable molecular distance geometry problem, the molecule conformation is resolved (Fig. 4(e)) with a resolution 0.3 nm.

\begin{figure*}
\centering
\includegraphics[width=2\columnwidth]{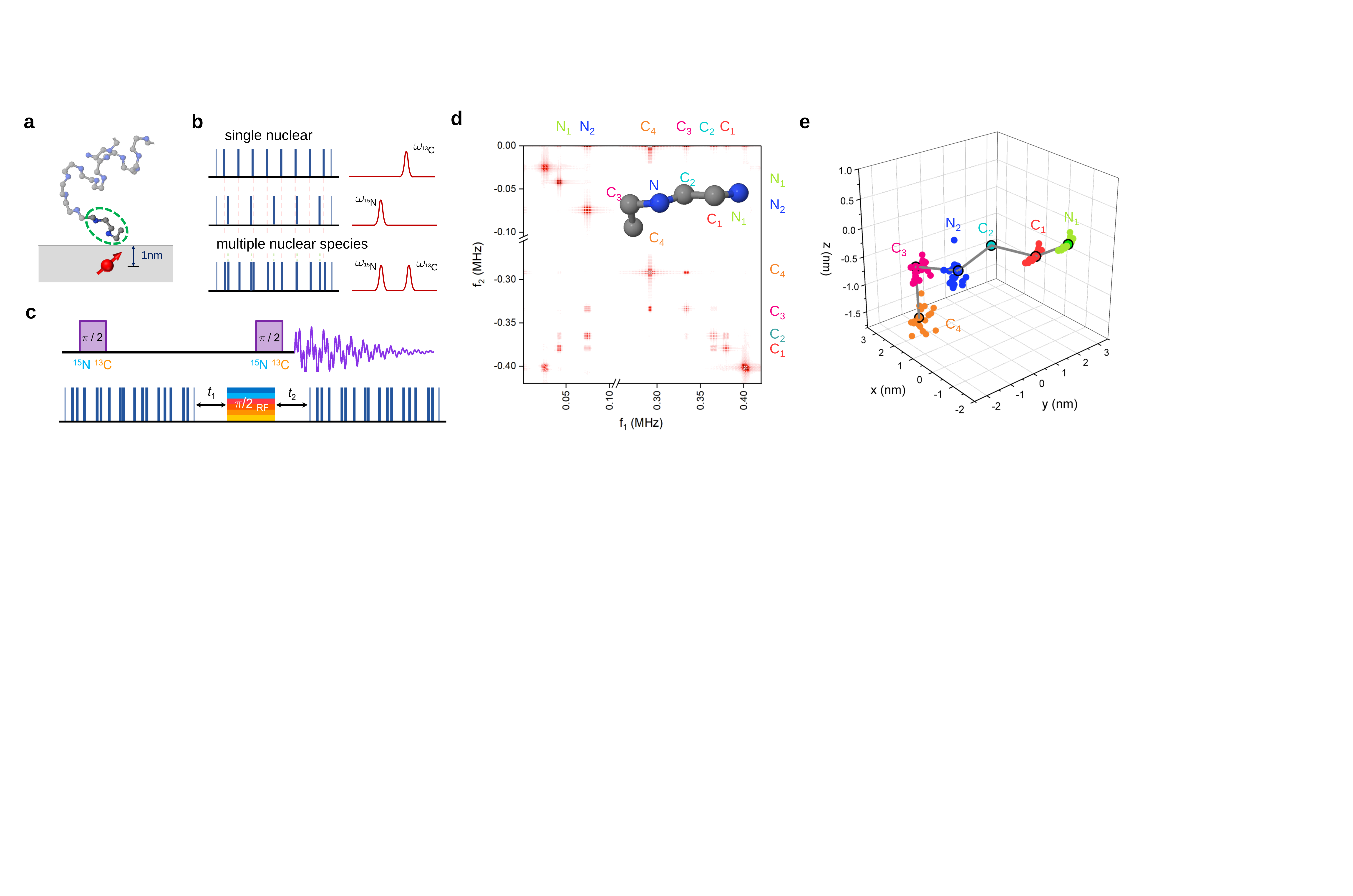}
\caption{Molecule structure analysis by 2D nanoscale NMR.
(a) The experimental scheme. A protein molecule placed 1nm from NV sensor. The structure of two amino acids labeled with $\cnuc$ and $^{15}\text{N}$ is detected with NV sensor. (b) Pulse sequence for homo- and hetero- NMR spectra. Upper and middle, periodical pulse sequence for different single nuclear species, the same as the dynamical decoupling sequence previously used. The pulse sequence filters out single nuclear species signal. Lower, nonperiodical pulse sequence probing two nuclear species at the same time. (c) Pulse sequence for homonuclear and heteronuclear magnetic resonance spectra. Similar to homonuclear spin two-dimensional NMR sequence, multi-nuclear species sequence works as initialization and correlation part, the middle RF $\pi/2$ pulse operates on all nuclear species. (d) The 2D nanoscale single molecule NMR simulated according to the pulse sequence in (c). (e) From the simulated single molecule two-dimensional NMR spectrum, the 3D geometry of protein structure is then reconstructed from the couplings with standard deviation 0.3 nm.
}\label{fig:mol}
\end{figure*}

Overall, the  2D nanoscale NMR spectrum of a single  cluster of two coupled nuclear spins is demonstrated by using a quantum sensor of the NV center in diamond. Equipped with a high sensitive NV sensor, the length of chemical bonds is resolved under 2D nanoscale NMR in diamond. Although the structure of a single cluster of two nuclear spins has previously been resolved \cite{shi_sensing_2014}, a 2D correlation spectroscopy is still necessary to resolve the structure of more complicated molecule or nanostructure. The 2D nanoscale NMR spectroscopy reported here, together with previous work on nanoscale NMR, can yield valuable structural information as opposed to bulk NMR, where the couplings typically hamper  structure analysis.  The spectral resolution can be improved by, for example, the Qdyne technique \cite{schmitt_submillihertz_2017,boss_quantum_2017} and  weak measurement readout \cite{cujia_watching_2018,pfender_high-resolution_2018}. The sensitivity can be improved by more efficient readout method \cite{shields_efficient_2015}. With further improvements, our 2D nanoscale NMR method has the potential to characterize the structure of a single molecule or conformational dynamics at single molecule level that is not accessible by conventional methods.

In the process of preparing the manuscript, a similar work just appears on arXiv \cite{abobeih_atomic-scale_2019}.

\section{Acknowledgments}
We thank Fedor Jelezko for helpful discussions. The authors at University of Science and Technology of China are supported by the National Key R$\&$D Program of China (Grant No.~2018YFA0306600, No.~2016YFA0502400, No.~2018YFF01012501, and No.~2017YFA0305000),
the National Natural Science Foundation of China (Grants No.~81788101, No.~91636217, No.~11761131011, No.~11722544, and No.~11775209),
the CAS (Grants No.~GJJSTD20170001, No.~QYZDY-SSW-SLH004, and No.~YIPA2015370),
Anhui Initiative in Quantum Information Technologies (Grant No.~AHY050000), the CEBioM, the Thousand-Young-Talent Program of China, and the Fundamental Research Funds for the Central Universities, the Innovative Program of Development Foundation of Hefei Center for Physical Science and Technology (Grants No.~2017FXCX005).

\section*{Supplementary materials}
Supplementary Text

Figs. S1 to S11



\end{document}